\documentclass[twocolumn,preprintnumbers,amsmath,amssymb]{revtex4-1}
\usepackage{graphicx}
\usepackage{dcolumn}
\usepackage{bm}

\begin{document}


\title{Interaction between single vacancies in graphene sheet: An \textit{ab initio} calculation}

\author{W.L. Scopel}
\email{wlscopel@if.uff.br}
\affiliation{Departamento de Ci\^encias Exatas, Universidade Federal Fluminense, Volta Redonda, RJ, 27255-250, Brazil}
\affiliation{Departamento de F\'{\i}sica, Universidade Federal do Esp\'irito Santo, Vit\'oria, ES, 29075-910, Brazil}
\author{Wendel S. Paz}
\email{wpascal1@gmail.com}
\affiliation{Departamento de F\'{\i}sica, Universidade Federal do Esp\'irito Santo, Vit\'oria, ES, 29075-910, Brazil}

\author{Jair C.C. Freitas}
\email{jairccfreitas@yahoo.com.br}
\affiliation{Departamento de F\'{\i}sica, Universidade Federal do Esp\'irito Santo, Vit\'oria, ES, 29075-910, Brazil}

\begin{abstract}

In order to investigate the interaction between single vacancies in a graphene sheet, we have used
spin-polarized density functional theory (DFT). Two distinct configurations were considered, either with the two vacancies located in the same sublattice or in different sublattices, and the effect of changing the separation between the vacancies was also studied. Our results show that the ground state of the system is indeed magnetic, but the presence of
the vacancies in the same sublattice or in different sublattices and the possible topological
configurations can lead to different contributions from the $\pi$  and $\sigma$ orbitals to magnetism.
On the other hand, our findings reveal that the net magnetic moment of the system with the two vacancies in the same
sublattice move towards the value of the magnetic moment per isolated
vacancy with the increase of the distance between the vacancies, which is ascribed to the different
contributions due to $\pi$ electrons. Moreover, it is also found that the local magnetic moments for
vacancies in the same sublattice are in parallel configuration, while they have different orientations when the vacancies are created in different sublattices.
So, our findings have clearly evidenced how difficult it would be to observe experimentally the emergence of magnetic order in graphene-based systems containing randomly created atomic vacancies, since the energy difference between cases of antiferromagnetic and ferromagnetic order decreases quickly with the increase in the distance separating each vacancy pair. 

\end{abstract}
 


\maketitle

\section{Introduction}

Graphene, a two-dimensional material with a single atomic layer of graphitic carbon,
has attracted great attention recently, due to its novel electronic properties and
potential device applications \cite{santosCPL2011,zouPRL2010,SongNanotech2010}. 
Ideal graphene sheet is nonmagnetic, then to use graphene for spintronic applications
the major challenge is to make graphene magnetic. In this sense, the experimental observation of magnetic properties in graphene and related materials \cite{OhldagPRL2007,LuoNanoLett2007, nair2012,freitas2014hyperfine,nair2013dual} has drawn huge interest, especially considering the applications of these materials in spintronics, quantum information processing and others.

Recently theoretical works \cite{yazyev2,ugeda,boukhvalovPRB2011,PalaciosPRB2012,PazSSC2013,felixPRB2014}
have revealed that the magnetic properties in graphene-based systems are associated with the occurrence
of defects such as atomic vacancies, substitutional and chemisorbed species. In particular,
vacancies in graphene sheet giving rise to magnetism depend on
 the density of defects\cite{singhJPCM2009}. Single-atom defects in graphene also lead to the occurrence of 
quasi-localized states near the Fermi level \cite{PazSSC2013}. The fact that quasi-localized states
lie at the Fermi level suggests that itinerant magnetism can be induced due to electron exchange
instability \cite {eduwardsJPCM2006}. 

Here, we report a systematic investigation of the effects of changing the spatial arrangement of two vacancies in a graphene sheet on its electronic and magnetic properties; the separation between the vacancies as well as their location (with respect to the distinct sites of the bipartite graphene lattice) were studied, while keeping the density of the defects fixed.
Two single vacancies were created in the same sublattice (A0A$i$) or in different sublattices (A0B$i$) with $i$=1, 2, 3 and 4 for different distance between the vacancies.
Spin-polarized DFT calculations with full structural relaxation were performed.
Our findings show that the ground state of the graphene sheet containing two vacancies is indeed magnetic, with a ferromagnetic-like coupling between the magnetic moments associated with the vacancies. However, depending on the creation of the vacancies on the same sublattice or in different sublattices and on the possible topological lattice distortions, different contributions of the $\pi$
 and $\sigma$ orbitals to magnetism are observed.

\section{Calculations}
In this work, we have investigated the interaction between carbon vacancies in a
 graphene monolayer, through first-principles calculations, based on the density 
functional theory (DFT)\cite{dft1,dft2}. The spin-polarized DFT calculations were 
performed using ultrasoft Vanderbilt pseudopotentials \cite{vanderbilt}, and a 
generalized gradient approximation (GGA) for the exchange-correlation potential
\cite{perdew}, as implemented in the VASP code \cite{kresse1,kresse2,kresse3}. 
In this calculation we have used a 9$\times$9 supercell with 160 carbon atoms and two atomic vacancies, which were created by removing carbon atom from an pristine graphene sheet. Herein, two different vacancy locations were considered: (a) vacancies in the A 
sublattice and (b) vacancies in the B sublattice. The lattice parameter for graphene 
obtained from structural optimization was 2.46 \AA. We have used a 
plane-wave-cutoff energy of 400 eV and a Monkhorst-Pack \cite{monkhorst} scheme 
with a 3$\times$3$\times$1 k-mesh for the Brillouin zone integration. In all calculations the 
lattice parameter was kept fixed at the calculated value, whereas the atoms were allowed 
to relax until the atomic forces were smaller than 0.025 eV/\AA.

\section{Results and Discussions}

\begin{figure}[!h]
\begin{center}
\includegraphics[width = 8.5 cm]{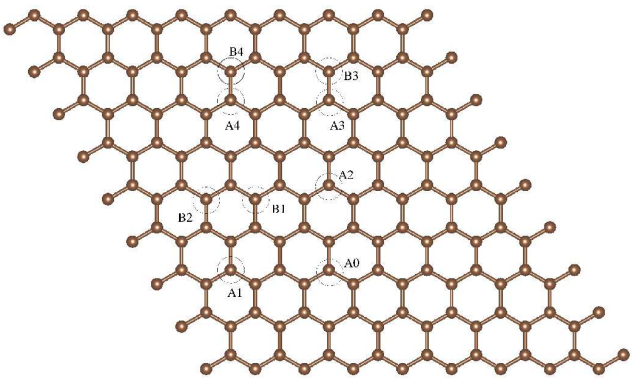}
\caption{Topological configurations of two vacancies on single graphene, 
where A{\textit i} and B{\textit i} indicate the position
 of the vacancies created on A and B sublattice, respectively. }
\label{structure}
\end{center}
\end{figure}

In order to study the effect of the interaction between two vacancies in the graphene 
sheet, we have considered several different configurations as shown in Figure \ref{structure}.
It is important to stress that the vacancy concentration was kept fixed in all calculations reported herein.

\begin{figure}[!h]
\begin{center}
\includegraphics[width=8.5cm]{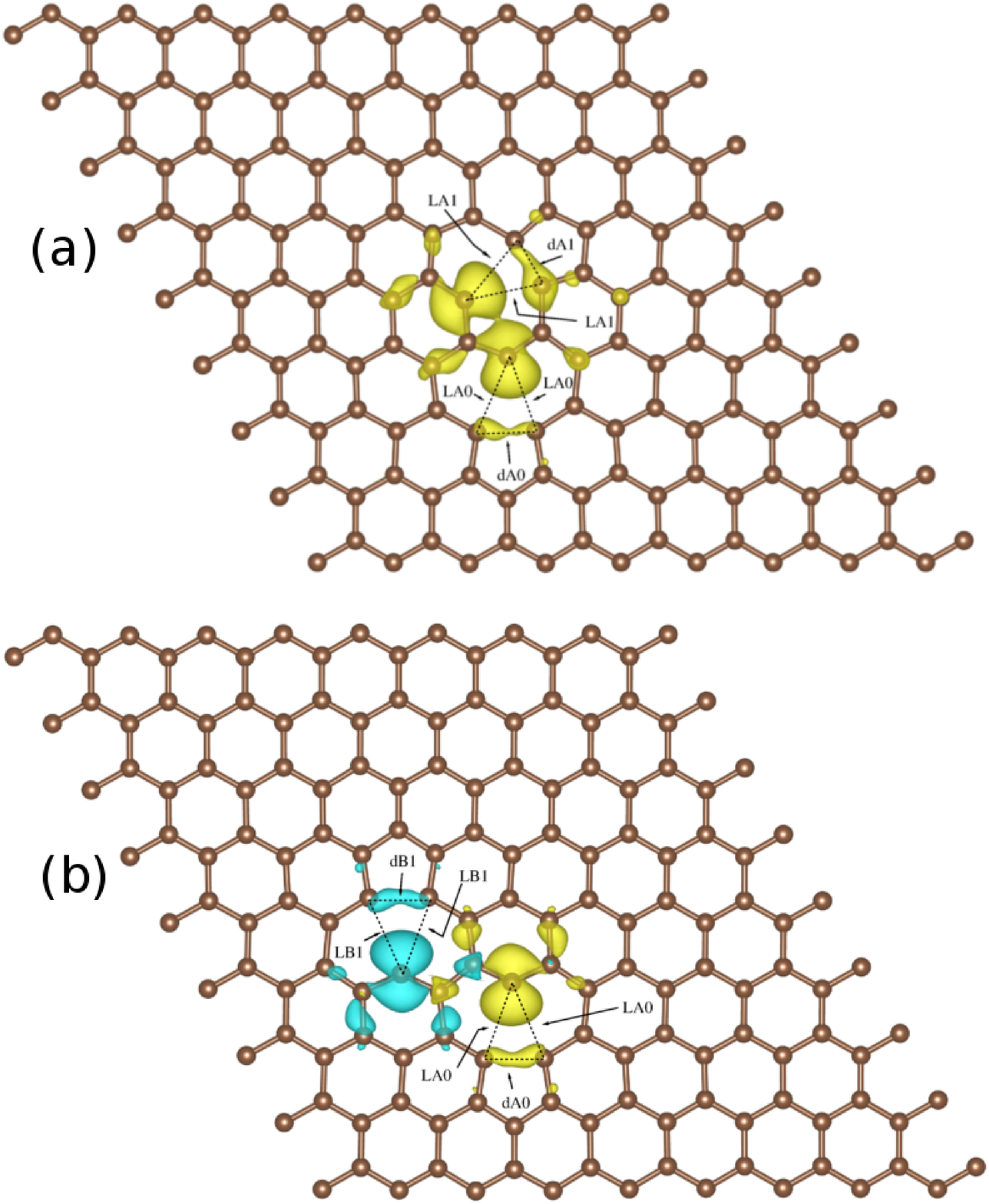}
\caption{Net spin density distribution for the ground states of the (a) A0A1 and (b) A0B1 configurations. The spin up and spin down states are represented by yellow and blue colors, respectively.}
\label{spinD}
\end{center}
\end{figure}

First, a single defect vacancy was created in A0 (as indicated in Figure \ref{structure}) and the structure was completely relaxed. The structural relaxation carried out for the entire structure yielded a planar Jahn Teller distorted carbon triangle and local magnetic moment around the vacancy, in accordance with previous work\cite{chenNanoscale2014}. For the carbon triangle (see Figure \ref{spinD}), we found two long bonds with a length of 2.55 \AA~ each and a short bond with a length of 1.98 \AA~ (see table \ref{triangle}), as compared with the value of 2.46 \AA~ for the
undistorted structure. The formation of this short bond points to a reconstruction of the two dangling bonds left after the removal of the carbon atom from the A0 site in the graphene sheet \cite{PazSSC2013}. The ground state of the relaxed system exhibited a magnetic moment 1.16 $\mu_B$, in agreement with previous reports \cite{PalaciosPRB2012}.
 
The formation energy of a single vacancy in the graphene sheet at the ground state of the system was calculated as

 \begin{equation}
 \label{formation}
 E_f=\frac{1}{n}(E_v-\frac{N-n}{N}E_{g}),
\end{equation}

where $N$ denotes the number of C atoms in the defect-free graphene sheet, $n$
represents the number of vacant C atoms (i.e., $n=1$ for a single vacancy) and $E_{g}$ and  $E_v$ are the total energies
 of the  defect-free and the vacancy-containing  graphene sheets, respectively. 
 The formation energy of the graphene sheet containing a single vacancy calculated according to
equation (\ref{formation}) was found to be 7.6 eV, which is in good agreement with
the experimental value of 7.0 eV \cite{mayer} and also with previous results of
DFT calculations \cite{dai,faccio2}.

However, when we created the second vacancy in the graphene sheet the formation energy was found to be 7.50 eV, which is a value somewhat smaller than the one corresponding to a single isolated vacancy \cite{PazSSC2013}. A Jahn-Teller distorted triangle was again observed around each vacancy (see Figure \ref{spinD}). The values of the bond lengths in these triangles for a graphene sheet with two
 vacancies separated by varying distances after a full relaxation of the whole topological configuration (see figure \ref{structure}) 
are given in Table \ref{triangle}, with the symbols d and L standing for the short bond and the two long bonds, respectively.

\begin{table}[h!]
\begin{center}
\caption{Bond lengths in the carbon triangles formed around each single vacancy in the relaxed graphene structure. The symbols dA$0$ and LA$0$ represent the short bond and the two long bonds in the triangle formed around the A0 site, whereas dX$i$ and LX$i$ represent the corresponding bond lengths in the triangle formed around the second vacancy (X can indicate an A or B site and $i$ ranges from 1 up to 4.)}
\label{triangle}
\begin{tabular}{c c c c c c c}
\hline 
\hline 
system & dA0(\AA) & \multicolumn{2}{c}{LA0(\AA)}  &dX$i$(\AA)&\multicolumn{2}{c}{LX$i$(\AA)}\\ 
\hline 
\hline 
A0 or B0& 1.98    & 2.55      &2.55     & 1.98    & 2.55      &2.55 \\ 
A0A1     & 1.73    & 2.54      &2.52     & 1.90    & 2.58      &2.65 \\ 
A0A2     & 1.79    & 2.52      &2.56     & 1.86    & 2.60      &2.68 \\ 
A0A3     & 1.87    & 2.54      &2.54     & 1.93    & 2.57      &2.57 \\ 
A0A4     & 1.89    & 2.55      &2.58     & 1.95    & 2.58      &2.65 \\  
A0B1     & 1.91    & 2.61      &2.64     & 1.91    & 2.61      &2.64 \\  
A0B2     & 1.84    & 2.53      &2.60     & 1.82    & 2.56      &2.57 \\  
A0B3     & 1.97    & 2.57      &2.59     & 1.95    & 2.55      &2.55 \\  
A0B4     & 2.08    & 2.57      &2.64     & 1.93    & 2.55      &2.57 \\
\hline 
\hline 
\end{tabular} 
\end{center}
\end{table}

From the data displayed in Table \ref{triangle}, one can note that the average lengths of the two long bonds are in general larger than 2.55 \AA~(value corresponding to a single isolated vacancy) for all configurations. Also, for systems with two vacancies, it can be verified that the length of the short bond in the carbon triangles increases with the increase of the distance between the vacancies; however, the average length of this short bond calculated from the lengths associated with each vacancy (dX$i$ and dA0) remains in most cases below the value corresponding to an isolated vacancy (1.98 \AA).

\begin{figure}[h]
\begin{center}
\includegraphics[width = 8.5 cm]{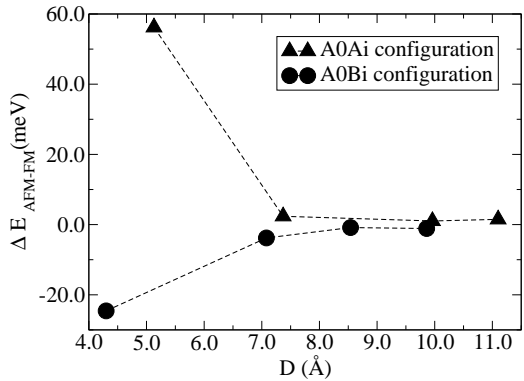}
\caption{Total energy difference between AFM and FM states ($\Delta E_{AFM-FM}$) as a function
 of the distance (D) between the single vacancies.}
\label{energy}
\end{center}
\end{figure}

In order to investigate the effect of the creation of the second vacancy on the ground state of the system and on its total magnetic moment, different configurations were considered in terms of the relative location of each vacancy (A0A$i$ or A0B$i$ configurations) and of the separation between the vacancies (different      $i$ values), as indicated in Figure \ref{structure}. 
In this sense, the total energy difference between states with distinct magnetic couplings involving the magnetic moments associated with each vacancy was determined. The values of the calculated total energy differences considering ferromagnetic (FM) and antiferromagnetic (AFM) coupling for all configurations studied herein are shown in Figure \ref{energy}. 

The results plotted in Figure \ref{energy} clearly show that the FM state is more energetically 
stable than the AFM state for the case of the two vacancies created on the same sublattice (A0A$i$ configuration). The ferromagnetic coupling between the magnetic moments associated with the two single vacancies can be seen in Figure \ref{spinD}a in the case of the ground state of the A0A1 topological configuration, where only spin up (yellow) electron states are observed around the vacancies. On the other hand, for vacancies created on different sublattices, the AFM state is more energetically
stable than the FM state, which follows closely Lieb's theorem for bipartite lattices \cite{LiebPRL1989}. 
For the ground state of the A0B1 configuration (see Figure \ref{spinD}b), one can observe the presence of contributions due to spin up (yellow) and spin down (blue) states on A and B sublattices, respectively, as expected for an antiferromagnetic coupling. These results are in good agreement with previous theoretical works\cite{yazyev2,helm,chenNanoscale2014},
where DFT calculations predict antiferromagnetic order in the case of vacancies on different sublattices and ferromagnetic order otherwise.
One can also note for both FM and AFM states that for all topological configurations the magnitude of the total energy difference
 $\vert \Delta E_{AFM-FM}\vert$ decreases with the increase of the distance 
between the single vacancies, becoming negligible for distances D larger than 7 \AA. This finding helps to explain the difficulty (or the impossibility, in some cases) to observe experimentally the occurrence of magnetic order in graphene-based systems, since the defect concentration needs to be considerably high so as to produce a significant energy imbalance between FM and AFM states. For instance, Nair et al. recently reported the occurrence of paramagnetic behavior down to liquid helium temperatures in graphene samples presenting high defect concentrations (close to $10^{20} g^{-1}$), with the average separation between magnetic moments estimated at 10 nm \cite{nair2012}. The present results clearly show that this spacing is about 10 times larger than the threshold above which the energy difference between AFM and FM couplings becomes negligible. Thus, one could expect that, in graphene-based materials with this range of vacancy concentration, any indication of magnetic order can be completely ruled out. For vacancies randomly created on A and B sublattices and separated by distances above ca. 10 \AA, the results of our calculations clearly indicate there will be no definite magnetic order in the system.     

\begin{figure}[!h]
\begin{center}
\includegraphics[width = 8.5 cm]{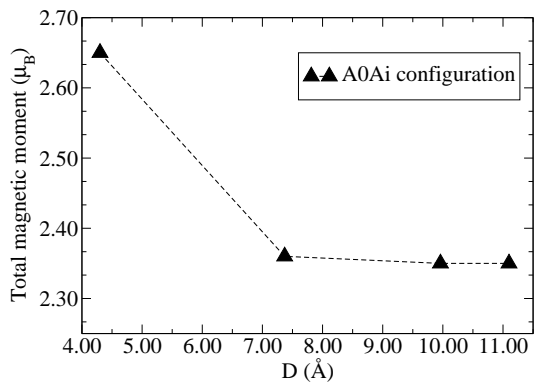}
\caption{Total magnetic moment as a function of distance (D) between two vacancies on same sublattice.}
\label{moment}
\end{center}
\end{figure}

\begin{figure}[!h]
\begin{center}
\includegraphics[width=8.5cm]{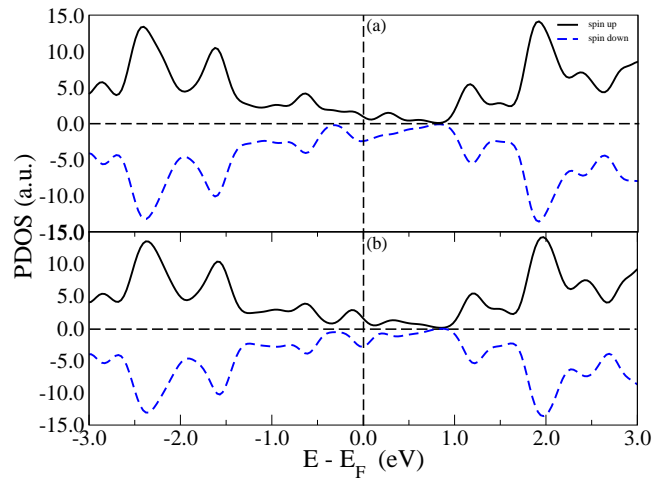}
\caption{ Projected density of states (PDOS) on p$_z$ orbital of the carbon atoms in a graphene sheet with two vacancies in A0A$i$ configuration:
(a) A0A1 (b) A0A4.}
\label{magmom}
\end{center}
\end{figure}

Furthermore, we explored the effect of changing the distance between the two vacancies on the net magnetic moment of the system, in the case of the A0A$i$ configurations (where the ground state corresponds to FM coupling). Figure \ref{moment} shows there is a reduction of the total magnetic moment as the distance between the vacancies grows. It is also important to note that the net magnetic moment approaches a value equal to twice the magnetic moment of an isolated vacancy when the vacancies become too far apart. Thus, as the long range interaction between the two vacancies becomes negligible, the magnetic moment per vacancy is similar to the one corresponding to a single isolated vacancy located at the same sublattice. Moreover, one can also observe that for D around 5 \AA~, the A0A$1$ configuration, the total magnetic moment is somewhat higher than the double of the total magnetic moment of the isolated vacancy, clearly showing the effect of the FM coupling between the magnetic moments in this case. For all topological configurations, one can see that local magnetic moments are induced around the vacancies (as illustrated in Figure \ref{spinD}a) and the total magnetic moment has contributions from $\sigma$ and $\pi$ states. These results are in general agreement with previous calculations conducted in similar systems \cite{faccio2,francis2015nonchalant}. However, it is important to stress that the maximum values of the total magnetic moment in all cases here described are well below the values recently reported by Francis et al. \cite{francis2015nonchalant}, where total magnetic moments of ca. 3 $\mu_B$ are found for two vacancies located at the same sublattice and separated by more than 6.0 \AA. In contrast, our results show that the largest magnetic moment, which is ca. 14 $\%$ above the value corresponding to twice the magnetic moment of a single isolated vacancy, is found only for two vacancies quite close to each other (separation D around 5 \AA~, corresponding to the A0A$1$ configuration).  
 It is clearly observed in Figure \ref{magmom}a, 
that there is a strong spin splitting of $\pi$ orbitals, especially in the A0A1 case, which is due to the interaction
between the neighboring vacancies. This leads to an increase of the $\pi$ orbital contribution to the total magnetic 
moment, making it larger as compared to twice the value corresponding to a single isolated vacancy. This effect of spin
splitting of $\pi$ orbitals is reduced with the increase of the
distance between the two vacancies  (see  Figure \ref{magmom}a and b), which is accompanied by the reduction of the net magnetic moment shown in Figure \ref{moment}. 

These results are in good agreement with what is known about the ground state of graphene-based systems with vacancy-induced magnetism, which is related to Stoner instability giving rise to ferromagnetism due to delocalized $\pi$ electrons and localized moments of $\sigma$ dangling bonds \cite{ozakiPRB2014}. Then, the ground state of a graphene sheet containing two vacancies on the same sub-lattice is spin-polarized with
net magnetic moment varying according to the topological configuration. Additionally, for defects
created on different sublattices, our results follow closely the prediction of Lieb’s theorem, which indicates
the imbalance between the two sublattices is the origin of magnetism in graphene,
leading to an antiferromagnetic ground state in these cases.

\section{Conclusion}

In summary, we have used spin-polarized DFT calculations to investigate the magnetic interaction between two atomic vacancies in a graphene sheet for different topological configurations, keeping fixed the defect concentration. In this study, we have considered 
vacancies created either on the same or on different sublattices for various distances between them. The general conclusion emerging from this analysis was that the ground state of the system corresponds to a ferromagnetic or antiferromagnetic coupling of the magnetic moments associated with the interacting vacancies in the case of vacancies created on the same sublattice or in different sublattices, respectively.
The results revealed that the net magnetic moment for
the case of two vacancies on different sublattices moves towards the double of the magnetic moment of the isolated
vacancy with the increase of the vacancy separation. This trend indicates that the vacancy-vacancy interaction becomes negligible for long distances and that the magnetic moment per vacancy is similar to the case of a single isolated vacancy.  However, for the configurations with a short distance between the two vacancies, the
 interaction between the magnetic moments associated with the vacancies  conduct to different effects on the total magnetic moment for each type of configuration. This effect is attributed to the magnitude of the spin splitting of $\pi$ states.
Finally, our results have clearly evidenced how difficult it would be to observe experimentally the emergence of magnetic order in graphene-based systems containing randomly created atomic vacancies, since the energy difference between cases of antiferromagnetic and ferromagnetic order decreases quickly with the increase in the distance separating each vacancy pair. 

\section*{Acknowledgments}

The authors acknowledge the support from Brazilian
 agencies CNPq and CAPES; we thank also to CENAPAD-SP for computer time.
 
\bibliography{WLScopel}

\end{document}